\documentclass[onecolumn,aps,preprint,amsmath,amssymb, showpacs,showkeys]{revtex4}
\usepackage{graphicx}
\usepackage{latexsym}
\usepackage{dcolumn}
\usepackage{inputenc}
\usepackage{xcolor}
 \usepackage{amsmath,empheq}
\usepackage{bm}

\begin{document}
\newcommand{\eq}{\begin{equation}}                                                                         
\newcommand{\eqe}{\end{equation}}

\title{Analytical investigation of time-dependent two-dimensional non-Newtonian boundary layer equations}

\author{Imre F. Barna$^{1}$,  G. Bogn\'ar$^{2}$,  L. M\'aty\'as$^{3}$ and K. Hricz\'o$^{2}$}
\address{ $^1$ Hungarian Research Network, Wigner Research Centre for Physics, 
\\ Konkoly-Thege Mikl\'os \'ut 29 - 33, 1121 Budapest, Hungary \\
$^2$ University of Miskolc, Miskolc-Egyetemv\'aros 3515, Hungary, \\ 
$^3$Department of Bioengineering, Faculty of Economics, Socio-Human Sciences and 
Engineering, Sapientia Hungarian University of Transylvania,  
Libert\u{a}tii sq. 1, 530104 Miercurea Ciuc, Romania} 
 
\date{\today}
%Email: barnai@sunserv.kfki.hu
\date{\today}
  
%\maketitle
%%%%%%%%%%%%%%%%%%%%%%%%%%%%%%%%%%%%%%%%%%%%%%%%%%%%%%%%%%%%%%%%%%%%%%%
\begin{abstract} In this study, five different time-dependent incompressible non-Newtonian boundary layer models in two dimensions are investigated, including external magnetic field effects. The power law, the Casson fluid, the Oldroyd-B model, the Walter fluid B model and the Williamson fluid are analyzed. For the first two models, analytical results are given for the velocity and pressure distributions, which can be expressed by different types of hypergeometric functions. Depending on the parameters involved in the analytical solutions of the nonlinear ordinary differential equation obtained by the similarity transformation, a very wide range of solution types is presented.  
  
\end{abstract}


%\draft

\maketitle
\section{Introduction}
The scientific field of classical fluid mechanics is vast and cannot be summarised in a few finite arbitrary books. 
If we restrict ourselves to non-Newtonian fluids or boundary layer flows, the relevant literature is still remarkably large. The fundamental physics of such fluid flows can be found in basic textbooks such as \cite{arista}, \cite{non-newt}, \cite{patel}  \cite{schli}.  
In our previous publication on time-dependent self-similar solutions of compressible and incompressible heated boundary layer equations \cite{barna-bound}, we collected the most relevant literature in the field, which we omit here. 
Some analytic results are also available for stationary non-Newtonian boundary layers, for example, \cite{saer}, \cite{bog}, \cite{bog2}, \cite{mak}. 

Today, there is a growing need to study the motion of non-Newtonian fluids, which are often found in industrial applications and in modelling many manufacturing processes (heat exchangers, pharmaceuticals, food, paper). All fluids for which the shear stress-shear rate relation cannot be described by Newton's law are called non-Newtonian fluids. They can be characterized by formulas of an extremely diverse nature. 
Without completeness, we will investigate five different non-Newtonian fluid models in our 
study these are the following:
% Power law
The first is the non-Newtonian power-law model describes the viscosity of time-independent flow behaviour for both shear thinning (or pseudoplastic) and shear thickening (or dilatant) cases (see  \cite{arista}, \cite{schli}). This model is used in vast engineering applications, for instance, lubricants,  polymers,  and slurries  \cite{yori}, this model is well suited for numerical and analytical studies of fluid flows \cite{bog}, \cite{bog2}. 
 
% Casson
The second can be directly derived from the usual Newtonian fluids.  
This is the widely used viscosity model for viscoplastic non-Newtonian behaviour and 
named the Casson fluid model. Fluids that behave like a solid when the shear stress is less than the yield stress applied to a liquid, and start to deform when the shear stress is greater than the yield stress \cite{casson}.

% Oldroyd
Our third model is the non-Newtonian Oldroyd-B fluid is used in the literature for the rheological characterization of a type of viscoelastic fluid. Oldroyd applied the principle that stresses in a continuous medium can only arise from deformations and cannot change if the material is only rotated \cite{old}. 
% Walters

The fourth is the Walters-B model which was proposed by Walters for viscoelastic fluids \cite{walt}. It is a generalisation of the Oldroyd-B model and is used to describe the behaviour of fluids that are mainly used in food and food processing technologies.

% the Williamson fluid 
The final model is the Williamson fluid model which can also be used to describe the viscoelastic shear thinning properties of non-Newtonian fluids \cite{wil}. In Williamson's fluid model, the effective viscosity should decrease infinitely as the shear rate increases, which is effectively infinite viscosity at rest (no fluid motion) and zero viscosity as the shear rate approaches infinity.

The actual existing literature on non-Newtonian boundary layers is enormous, therefore we cannot give a general overview just mention some relevant recent 
publications in the right place in this study. 

% MHD
To analyse the variation in flow characteristics of a fluid in a magnetic field, an additional magnetic term is added to the momentum equation of the boundary layer equation system, so that the boundary layer equation system can describe the magnetic effects.

The aim of this paper is to provide analytic results for the stationary equations of the non-Newtonian fluid flows in a magnetic field. The above given five types of non-Newtonian behaviour are examined. The governing equations are transformed into a system of ordinary differential equations by a similarity transformation. We analyze the cases in which the fluid flow rate functions can be analytically given by the method used. In these two cases, we give the velocity distributions in two space dimensions and point out the effect of each parameter.
%%%%%%%%%%%%%
\section{Mathematical models}
\subsection{Power-law viscosity}
\begin{figure}  
\scalebox{0.6}{
\rotatebox{0}{\includegraphics{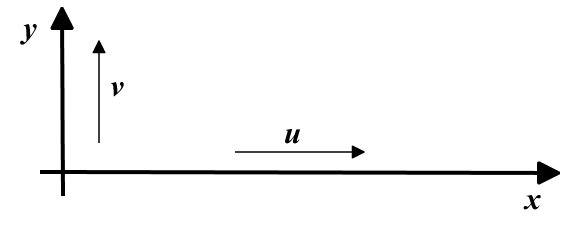}}}
\caption{Defining the directions and the velocity components of the investigated system.}
\label{nullas}      
\end{figure}

For power-law viscosity, our partial differential equation system systems reads as follows: 
%the momentum  equation  under investigation is:
%%%%%%%%%%%%%%%%%%%%%%%%%%%%%%%%%
\begin{eqnarray}
\frac{\partial u}{\partial x} + \frac{\partial v}{\partial y} &=& 0,  \label{pde1} \\ 
\rho\left( \frac{\partial u}{\partial  t} +   u\frac{\partial u}{\partial x} + v\frac{\partial u}{\partial y} \right) &=& - \frac{\partial p}{\partial  x} 
+  m \frac{ \partial   }{\partial y} \left(\left|\frac{\partial u}{\partial y}\right|^{n-1}  \cdot \frac{\partial u}{\partial y} \right) + \sigma B_0^2 u, \label{pde2} \\ 
\frac{\partial p}{\partial  y} &=&0, 
\label{pde3} 
\end{eqnarray}
where the dynamical variables are the velocities components $u(x,y,t), v(x,y,t)$, the fluid pressure $p(x,y,t)$, 
% $B$ B = B (t)= B_0 / \sqrt{t}$ with 
$B_0 = const $ is the magnetic 
the magnetic induction, and $\sigma$ is the electrical conductivity.  
Additional physical parameters are  $\rho, m, n$
the fluid density, consistency parameter, and the power-law index, respectively. 
To fix the geometrical relations in our system consider Fig.(\ref{nullas}).

The temporal decay of the strong solutions of the power-law type of non-Newtonian fluids for variable power-law index was mathematically proven by Ko \cite{power1}. 
Flow reversal effects in an expanding channel for this type of viscous fluid were analyzed by 
\cite{power2}. A non-iterative transformation method to an extended Blasius problem describing a 2D laminar boundary-layer with power-law viscosity for non-Newtonian fluids was developed by Fazio \cite{power3}. 
The stationary power-law equations were solved by Pater {\it{et al.}} \cite{patel} by the one-parameter deductive group theory
technique. (We have to mention at this point that the stationary boundary layer equations can be reduced to the third-order non-linear differential equation which is called the Blasius equation having an extensive literature which we skip now.)  

To obtain the solutions to the system (\ref{pde1})-(\ref{pde3}), we apply the following self-similar Ansatz for $u, v$ and $p$:  
\begin{eqnarray}
u(x,y,t) = t^{-\alpha} f(\eta), \hspace*{1cm}
v(x,y,t)  = t^{-\delta} g(\eta),       \hspace*{1cm}
p(x,y,t) =  t^{-\gamma}h(\eta),
\label{ansatz}
\end{eqnarray}
with the argument $\eta = \frac{x+y}{t^{\beta}}$ of the shape functions.  
All the exponents $\alpha,\beta,\gamma,\delta $ are real numbers. Solutions with integer exponents are called 
self-similar solutions of the first kind, non-integer exponents generate self-similar solutions of the second kind \cite{sedov}.

It is necessary to remark that, the heat conduction mechanism to (\ref{pde1}) - (\ref{pde3}) the boundary layer equations can exclude the self-similarity using transformation (\ref{ansatz}) one gets contradiction among self-similar exponents.

The shape functions $f, g$ and $h$ could be any arbitrary continuous functions with existing first and second continuous derivatives and will be derived later on.  
The physical and geometrical interpretation of the Ansatz were exhaustively analyzed in  former publications \cite{imre1,imre3}, therefore we skip it here. 

The main points are, that $ \alpha, \delta, \gamma$ are responsible for the rate 
of decay and $\beta$ is for the rate of spreading of the corresponding dynamical variable for positive exponents. 
Negative exponents mean physically irrelevant cases, blowing up and contracting solutions.  
The numerical values of the exponents are considered as follows:  
\eq
\alpha = \delta = n/2, \hspace*{1cm} \beta = 1 - \alpha = 1- n/2, \hspace{1cm}  \gamma = n.   \label{five}
\eqe 
Exponents with numerical values of one-half mean the regular Fourier heat conduction 
(or Fick's diffusion) process. 
One-half values for the exponent of the velocity components and unit value exponent for the pressure decay are usual for 
the incompressible Navier-Stokes equation \cite{imre4}.  Note, that the value of  $\gamma$ is responsible for the decay of the pressure field. 
%%%%%%%%%%%%%%%%%%%%%%%%%%
°% fig nullas 
%%%%%%%%%%%%%%%%%%%%%%%%
Applying (\ref{five}), the derived ordinary differential equation (ODE) system reads as
\begin{equation}
f' + g' =  0, 
\label{six} 
\end{equation}
\begin{equation}
\rho \left(- \frac{n}{2} f - \left[ 1- \frac{n}{2}\right] \eta f' + ff' + gf'  \right)  = - h'  + mnf''|f'|^{n-1} + \sigma B_0^2 f,  
\label{seven}
\end{equation}
\begin{equation}
h' = 0,
\label{eight}
\end{equation}
where the prime means derivation concerning variable $\eta$.  
Equations (\ref{six}) and (\ref{eight}) can be integrated, to get $f + g = c_1$ and   $ h = c_2$. 
However, the dynamic variable under consideration is the velocity component $u$, which is $f$.   
From (\ref{seven}) the derived ODE is the following: 
\eq
\frac{mn}{\rho} f''  |f'|^{n-1}  + \left(\left[ 1- \frac{n}{2}\right] \eta - c_1\right) f'  + K f = 0, \label{nine}
\eqe
where $K = (\rho n/2 + \sigma B_0^2 ).$ The constant $c_1$ can be determined from possible boundary conditions applied to the fluid flow problem.
%%%%%%%%%%%%%%%%%%%%%%%%%%%%%%%%%%%%%%%%
%\section{Solutions}

Here, we give analytic and numeric solutions to the flow (\ref{six})-(\ref{eight}).

i.) First we consider the case $ B_0=0 $. Equation (\ref{nine}) has analytic solutions only for the cases of $n = 1, 0$ and $-1$.  
 
\begin{figure}  
\scalebox{0.8}{
\rotatebox{0}{\includegraphics{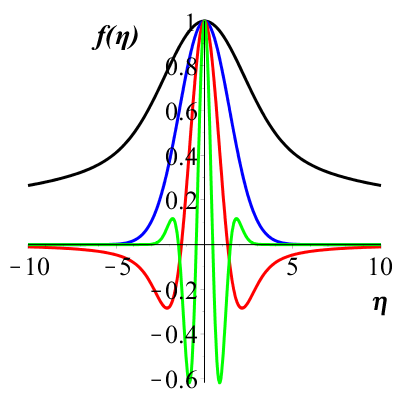}}}
\caption{Four different velocity shape functions for Eq. (\ref{v1a}). 
The black, blue, red and green curves 
are for $\rho = 0.5, 1, 2$ and $5$, respectively. 
All additional parameters are the same for all four curves ($C_1 =1, C_2 = 0, c_1 = 0$ and $m = 2$ ).}
\label{egyes}      
\end{figure}

\begin{figure}  
\scalebox{0.6}{
\rotatebox{0}{\includegraphics{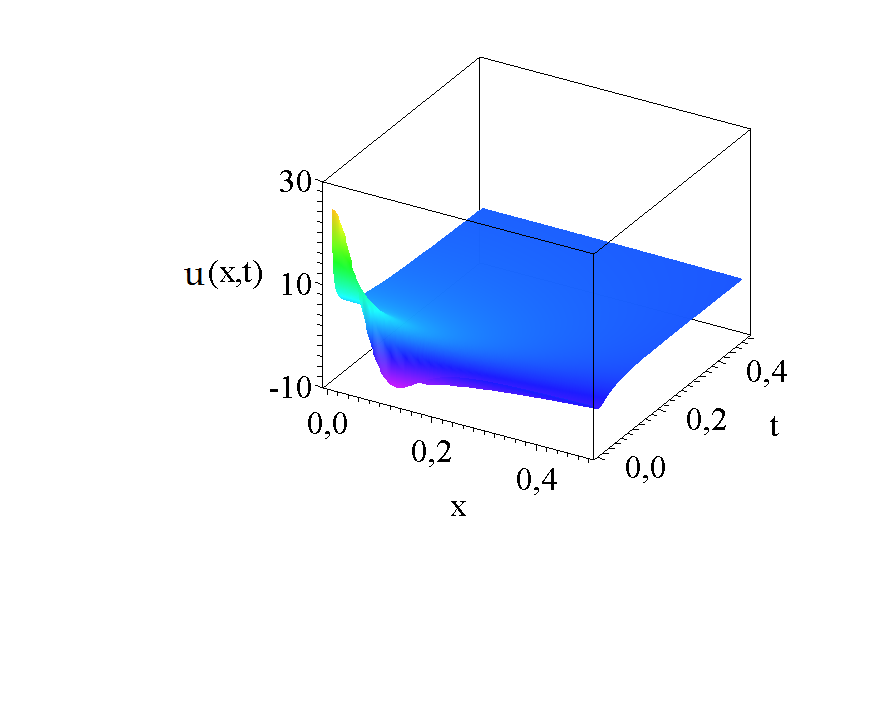}}}
\caption{The velocity distribution function projection for $y = 0$ for 
the density of $\rho = 2$. All other parameters are the same as below.}
\label{kettes}      
\end{figure}

For the Newtonian fluid case, when $n = 1$, and  other parameters ($c_1,m,\rho$) are  general, the solution reads as
\eq
f(\eta) =   C_1M\left(\frac{\rho}{2}, \frac{1}{2}, -\frac{\rho(\eta-2c_1)^2}
{4m}  \right) + C_2 U\left(\frac{\rho}{2}, \frac{1}{2}, -\frac{\rho(\eta-2c_1)^2}
{4m}  \right), 
\label{v1a}
\eqe
where $M(,,)$ and $U(,,)$ are the Kummer M and U functions \cite{NIST} 
and $C_1$ and $C_2$ stand for the usual integration constants.
The parameter $c_1$ just shifts the maxima of the shape function and $m$ the consistency parameter 
(which is positive) just changes the full width at half the maximum of the function. The larger the numerical value of $m$ the broader the shape function. The third parameter, the density of the fluid  
(which must be also positive) is the most relevant parameter of the flow which of course, meets our physical intuition. 
Figure (\ref{egyes}) shows four different velocity shape functions for different fluid 
densities. It is important to mention that for density $0 < \rho  $  
the shape function has a global maxima in the origin and a decay to zero, 
(larger densities mean quicker decay.)
however for $1 < \rho  $ the shape function shows additional oscillations as well.  
The final velocity field has the asymptotic form of 
$u(x,y=0,t) = t^{-\frac{1}{2}}f\left(\frac{x}{t^{1/2}}\right)$. 
For completeness,  we give the entire formula for $u$ as well:
\eq
u(x,y,t) = \frac{1}{t^{1/2}} \left(  C_1M\left[\frac{\rho}{2}, \frac{1}{2}, -\frac{\rho \left\{ \frac{(x+y)}{t^{1/2}} -2c_1 \right\}^2}{4m}  \right]  + C_2U\left[\frac{\rho}{2}, \frac{1}{2}, -\frac{\rho \left\{ \frac{(x+y)}{t^{1/2}} -2c_1 \right\}^2}{4m}  \right]   \right) .
\label{v1}
\eqe
We remark that this case was investigated in \cite{barna-bound}. Figure (\ref{kettes}) shows the velocity field of $v(x,y=0,t)$ for a parameter set. The slight oscillation and the quick decay are clear to see.  
It is interesting to mention here, that during  our investigations using  the self-similar Ansatz, we regularly get solutions which contain 
Kummer's function for hydrodynamic
processes like the B\'enard-Rayleigh convection \cite{imre1} 
or even for regular diffusion equation \cite{laci_roman}. 
%%%%%%%%%%%%%%%%%%%%%%%%%%%%%%%%%%%%%%
%%%%%%%%%%%%%%%%%%%%%%%%%%%%%%%%%%%%%%%%%%%%%%%%
%%%%%%%%%%%%%%%%%%%%%%%%%%%%%%%%%%%%%%%%%%%%%%%%
For  the case $ n= K = 0$  the solution is a trivial constant $f = c_2$. 

There are analytic solutions available for $n = -1$ unfortunately, not 
for the general case where all the other three parameters $(c_1,\rho,m)$ are completely free. 
The constant $c_1$ should be equal to zero and the ratio of $\rho/m$ should 
have some special values. The existing possibilities are not so many. 

\begin{figure}  
\scalebox{0.5}{
\rotatebox{0}{\includegraphics{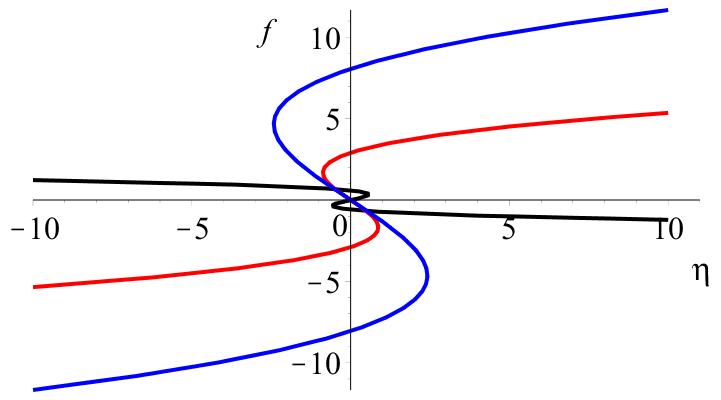}}}
\caption{The implicit velocity shape functions of Eq. (\ref{f3}) for three different parameter sets  $(C_1, \tilde{C}_2, m)$. 
The black, red and blue curves are for 
 $(1, 4, 0.4)$, $(2.4, 0.3, 1.3)$  and $(2.4, 0.3,10)$ .}
\label{harmas}      
\end{figure}

In the following we present some solutions which came from our numerical experimental experiences: \\
For $\rho = 1$ and for arbitrary integer or rational  m  
we get the next implicit relation for the solutions:  
\eq
 ln(\eta) - ln(f[\eta]) - ln\left(C_1 
U\left[-1,\frac{3}{2},\frac{1}{4m}f\{\eta\}^2 \right]  + 
M\left[-1, \frac{3}{2},\frac{1}{4m}f\{\eta\}^2 \right]\right) - C_2 = 0.    
\label{v2}
\eqe
If the first parameters of the Kummer's functions are non-positive integers 
the series becomes finite 
\begin{eqnarray}
M\left[-1,\frac{3}{2},\frac{1}{4m}f\{\eta\}^2 \right] = 1 - \frac{2}{3m} f(\eta)^2, \nonumber \\ 
U\left[-1,\frac{3}{2},\frac{1}{4m}f\{\eta\}^2 \right] = - \frac{3}{2} + \frac{1}{m} f(\eta)^2,
\end{eqnarray}
considering that $C_2 > 0$ and $\tilde{C_2} = ln( {C_2} )$ after some trivial algebraic steps we get:  
\eq
  \left( \frac{\tilde{C}_2 C_1}{m} - \frac{ 2\tilde{C}_2}{m} \right) f(\eta)^3 + 
 \left( \tilde{C}_2 - \frac{3 \tilde{C}_2 C_1}{2} \right)f(\eta) - \eta = 0. 
 \label{f3}
\eqe
The velocity shape functions can be seen in figure 
\ref{harmas}.
%%%%%%%%%%%%%%%%%%
% Fig 3 
%%%%%%%%%%%%%%%%%%%%%%%%%%%%%%

One can see on Eq. (\ref{f3}), that if $\eta=0$, than $f(\eta=0)=0$ is a solution of the equation. Further  possibilities gives the following rearrangement 
\eq
 \left[
  \left( \frac{\tilde{C}_2 C_1}{m} - \frac{ 2\tilde{C}_2}{m} \right) f(\eta)^2 + 
 \left( \tilde{C}_2 - \frac{3 \tilde{C}_2 C_1}{2} \right) \right] 
 f(\eta) = 0. 
 \label{f3-eta=zero}
\eqe
This yields
\eq 
f(\eta=0) = \pm 
\sqrt{\left(\frac{3 \tilde{C}_2 C_1}{2} - \tilde{C}_2\right)
/ \left( \frac{\tilde{C}_2 C_1}{m} - \frac{ 2\tilde{C}_2}{m} \right) }. 
\eqe
It shows that for $\eta=0$ there is one or there are three corresponding values to function $f(\eta)$. 
In the following we try to find the constraints on the 
parameters which separate the two cases. 
If we consider the inverse function $f(\eta)=y$, we have
\eq 
\eta = \left( \frac{\tilde{C}_2 C_1}{m} - \frac{ 2\tilde{C}_2}{m} \right) \cdot y^3 + 
 \left( \tilde{C}_2 - \frac{3 \tilde{C}_2 C_1}{2} \right)
 \cdot y.
\eqe 
The derivative of this function is 
\eq
\frac{d\eta}{dy} =  3 \left( \frac{\tilde{C}_2 C_1}{m} - \frac{ 2\tilde{C}_2}{m} \right) \cdot y^2  + 
 \left( \tilde{C}_2 - \frac{3 \tilde{C}_2 C_1}{2} \right) .
 \label{inv-derivative}
\eqe 
If this expression has always the same sign, the function is monotonous, and then there is just one single root 
$\eta(y=0)=0$ (i.e. $f(\eta=0)=0$). 
If this derivative may change the sign, then multiple roots 
are possible. 
The inverse function is not monotonous and may have three roots if there are real roots of the equation of derivative $d\eta/dy=0$. This means the following condition 
\eq
\frac{ 
\left(\frac{3 \tilde{C}_2 C_1}{2} - \tilde{C}_2\right)
}{ \left( \frac{\tilde{C}_2 C_1}{m} - \frac{ 2\tilde{C}_2}{m} \right)
} > 0 .
\eqe 

\begin{figure}  
\scalebox{0.90}{
\rotatebox{0}{\includegraphics{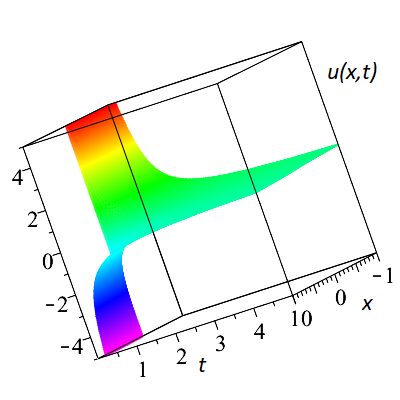}}} 
\scalebox{0.9}{
\rotatebox{0}{\includegraphics{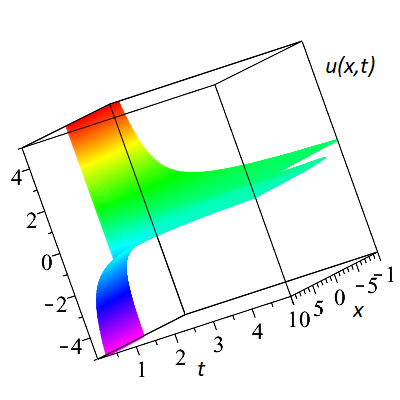}}} \\ 
 a)   \hspace*{3cm}   b)  
\caption{The implicit velocity distribution $u(x,t)$ evaluated from Eq. (\ref{f4}) for two different parameter sets     
 a) figure is for  $(1, 4, 0.4)$ and  b) figure 
  is for  $(0,2.5,0.7)$,     respectively.}
\label{negyes}      
\end{figure}

Figure (\ref{harmas}) shows two implicit solutions of Eq. (\ref{f3}) for three different parameter 
sets. Note, that Eq.(\ref{f3}) is a third-order equation we may get multi-valued solutions. 
If the parameter $m$ is much smaller than the two integration 
constants then single-valued solutions emerge as well. 
Whether the multi-valued solutions have any physical significance is not yet clear, but 
it is thought that this property may indicate the existence of finite oscillations or eddies. 
Figure (\ref{negyes}) shows the projected velocity fields $(y=0)$
for single and double-valued case. (For a better visibility we present 
the solution function from an unusual perspective.)
The decrease of the velocity field with time is clearly visible. 
For the sake of completeness, we give the final form of the velocity field, which reads: 
\eq
  \left( \frac{\tilde{C}_2 C_1}{m} - \frac{ 2\tilde{C}_2}{m} \right) t^{3/2}f \left(\frac{(x+y)}{t^{3/2}} \right)^3 + 
 \left( \tilde{C}_2 - \frac{3 \tilde{C}_2 C_1}{2} \right) t^{1/2}f \left(\frac{(x+y)}{t^{3/2}} \right)  - \frac{(x+y)}{t^{3/2}} = 0. 
 \label{f4}
\eqe

%%%%%%%%%%%%%%%%
% Fig 4 
%%%%%%%%%%%%%%%%

\begin{figure}  
\scalebox{0.5}{
\rotatebox{0}{\includegraphics{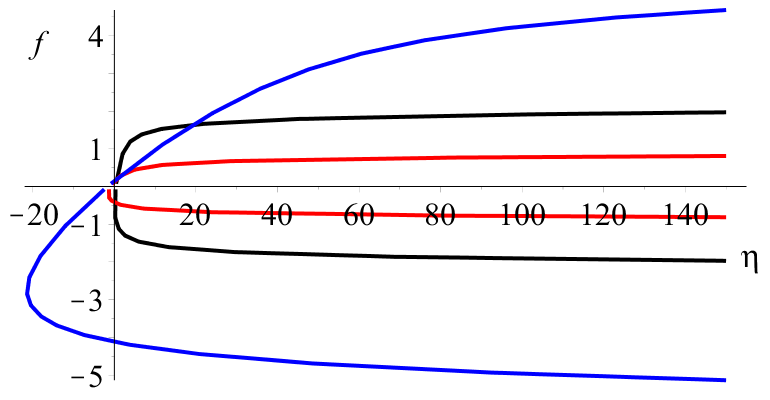}}}
\caption{The implicit velocity shape functions of Eq.(\ref{f4}) for three different parameter sets 
$(C_1, \tilde{C}_2, c_1, m)$ with $\rho =3$. 
The black, red and blue curves are for 
 $(1, 1, 1, 0, 1)$, $(3, 5, 0, 0, 0.2)$  and $(13, 11, 4, 10)$, respectively}. 
\label{otos}      
\end{figure}

\begin{figure}  
\scalebox{0.9}{
\rotatebox{0}{\includegraphics{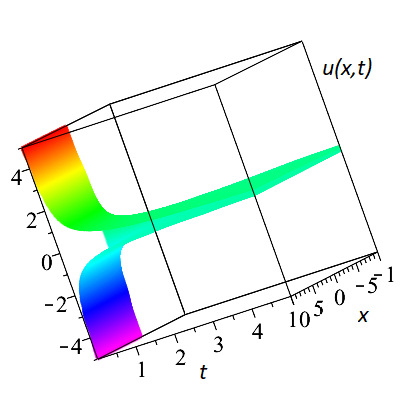}}}
\caption{The implicit velocity distribution field when  the parameters are $\rho =3$, 
$(C_1, \tilde{C}_2, c_1,m)$ and $(3, 5, 0, 0.2)$ with $\rho =3$.} 
\label{hetes_uj}      
\end{figure}

There are analytic solutions available for $\rho = 3$ and for arbitrary real $m$ and $c_1$  in the implicit 
form of: 
\eq
 \frac{\eta}{f(\eta)} + \frac{C_1 e^{\frac{9 f(\eta)^2 }{4m }}}{3f(\eta)} 
- \frac{ C_1 \cdot erf \left( \frac{3}{2} \sqrt{-\frac{1}{m} } f[\eta] \right)  }{2 m\sqrt{-\frac{1}{m \pi}}} - \frac{2c_1 }{3f(\eta)} -  C_2 = 0.  
\label{f4}
\eqe
Equation (\ref{f4}) clearly shows, that for $C_1 = 0$, we get the trivial linear function as a solution. 
Figure (\ref{otos}) presents the solutions of Eq. 
(\ref{f4}) for two parameter sets. 
 Because of the quadratic appearance of $f(\eta)$, we obtain dual-valued curves which are 
 not strictly speaking functions. The result functions increase strongly (or decrease in the case of a 
negative branch) very close to the origin and for larger arguments on a nearly 
horizontal plateau. The general structure of the solution is 
quite stable, even a tenfold change in the parameters does not change the overall feature of the derived curve.   
To have a better overview of the properties of the solution, 
Figure (\ref{hetes_uj}) presents the usual velocity projection of 
$u(x,y=0,t)$ for a given parameter set. 
The physically relevant rapid decay over time is again clearly visible. \\  
%%%%%%%%%%%%%%%%%%
% Fig 5 
%%%%%%%%%%%%%%%%%%%
% Fig 6 
%%%%%%%%%%%%%%%

ii.) In the second part of our analysis, we investigate the solutions when the magnetic 
induction is not equal to zero ($B_0 \ne 0$). Even now four different cases exist 
$ n = 2, 1, \ 0$ and $-1$. We have to examine them one by one. 

\begin{figure}{  
\rotatebox{0.8}{\includegraphics{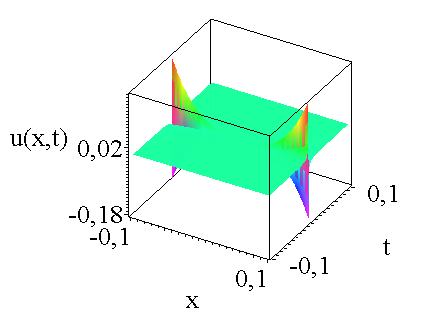}}}
\caption{The projection of Eq. (\ref{n2u}) $u(x,y=0,t)$ for the 
parameter set of 
$(\rho,m,\sigma,B_0,c_1)$ equal to $(1, 0.1, 1, 0)$, respectively.}
\label{n2}
\end{figure}

First, we consider the dilatant case  $n = 2$.  Equation (\ref{nine}) is reduced to
\eq
\frac{2m}{\rho} f''  f' + K f = 0. \label{nine_n2}
\eqe
multiplying by $f'$ we get a total derivative which 
can be integrated  giving us the ODE of: 
\eq
\frac{2m}{3\rho} f'^3 + \frac{1}{2}(\rho + \sigma B_0^2)f" + c_0 = 0.   
\eqe
First considering the $c_0 = 0$ simpler case,  
the ODE can be directly integrated and gives multiple solutions:
there is a trivial one of $f = 0$ there are two complex conjugated 
solutions which we skip and a relevant real one. 
After some algebraic manipulation, it reads as: 
\eq
f = -\frac{6\rho(\rho + \sigma B_0^2)(\eta - c_1)^3}{216 \cdot m}, 
\eqe
which is a third-order parabola in the variable of $\eta$. 
The final velocity distribution is 
\eq
u(x,y,t) = -\frac{6\rho(\rho + \sigma B_0^2)(x+y - c_1)^3}{216 \cdot m \cdot t}, 
\label{n2u}
\eqe
which is divergent at $t = 0$ for large spacial coordinates and has  
quick $t^{-1}$ decay in time in general. 
Figure (\ref{n2}) shows the graph of Eq. (\ref{n2u}) for a given parameter set.  

%%%%%%%%
% Fig n = 2 case 
%%%%%%%%%%%%%%

For $c_0 \ne 0$, there is a formal implicit solution containing an integral:  
\eq
\eta - \int^{f(\eta)} \frac{2m}{[-6 \rho m^2 (a^2{\rho + a^2\sigma B_0^2 +2c_0 })  ]^{1/3} }da - c_1 = 0.
\eqe
Unfortunately, for general arbitrary parameters $(\rho,m,\sigma,B_0^2, c_0)$ it cannot be evaluated. 

\begin{figure}  
\scalebox{0.4}{
\rotatebox{0}{\includegraphics{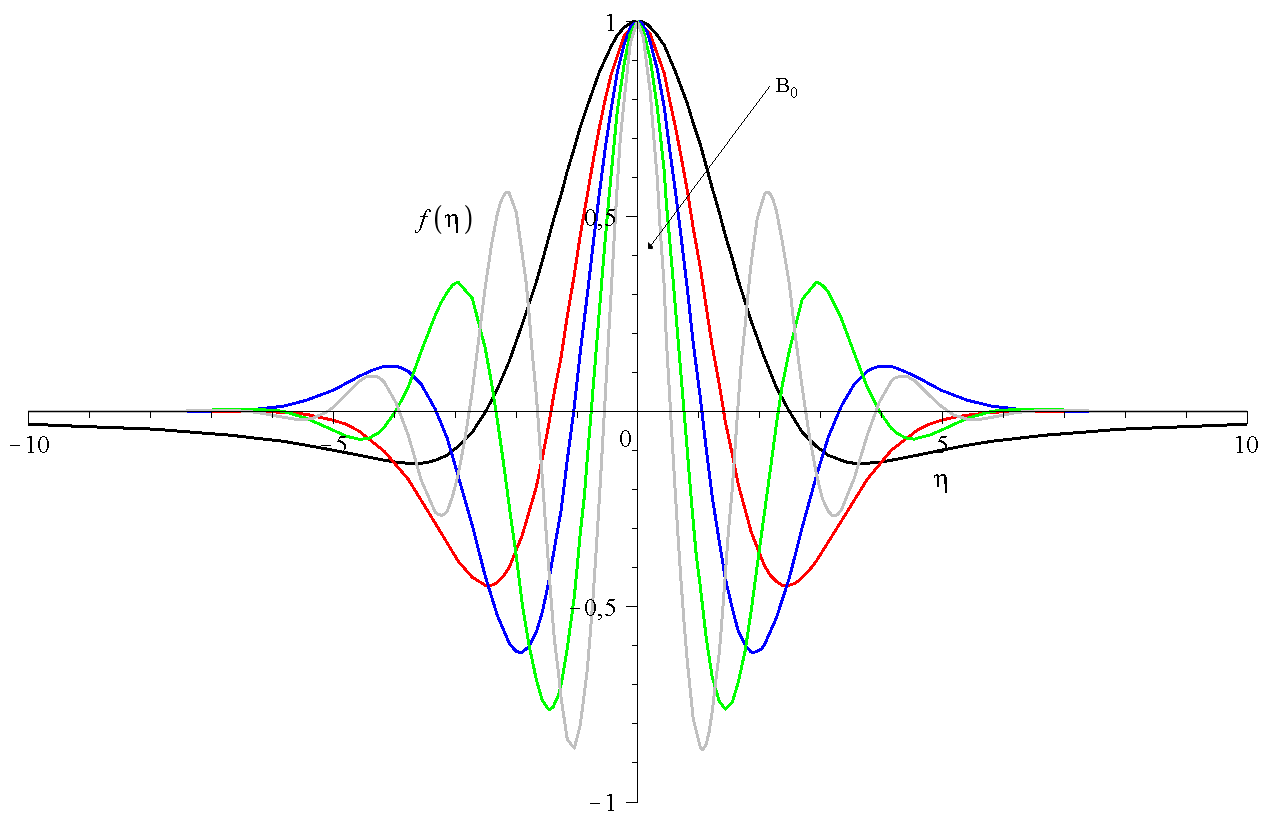}}}
\caption{Five shape functions of Eq. (\ref{f6}) for different parameter sets. 
The effect of the magnetic induction is investigated. 
The black, red, blue  and green lines are for $(C_1, C_2, c_1, \rho, m,  \sigma, B_0^2) $ with 
numerical values of $  (1,0,0,1,1,1,0.1)$,  $  (1,0,0,1,1,1,0.5)$,  
 $  (1,0,0,1,1,1,1)$,  $  (1,0,0,1,1,1,2)$, and  $  (1,0,0,1,1,1,4)$,
respectively}. 
\label{valamelyik}      
\end{figure}

The second case is $n = 1,$  the Newtonian fluid case:
\begin{eqnarray}
f(\eta) =  C_1 \cdot  
M\left(\frac{\rho}{2} + \sigma B_0^2, \frac{1}{2}, -\frac{\rho [ \eta-2c_1]^2}{4m}  \right) + 
 C_2 \cdot 
 U\left(\frac{\rho}{2} + \sigma B_0^2, \frac{1}{2}, -\frac{\rho [ 
\eta-2c_1]^2}{4m}  \right). 
\label{f6}
\end{eqnarray}
Note, the interesting feature that both the density and the magnetic induction give 
contributions into the first parameter of Kummer's functions.   
The density $\rho$ and the square of the magnetic induction 
$B_0^2$ are always positive the electric conductivity $\sigma$ is also positive 
for regular materials. (With the need for completeness we have to note that there are 
so-called metamaterials which can have negative electrical conductivity but we skip 
that case now. More on materials can be found in \cite{meta}. For such media the first parameter of Kummer's function 
would be negative which would mean divergent velocity fields, which 
contradicts energy conservation.)  
For regular materials, the external magnetic field enhances the first 
positive parameter of Kummer's functions which means more oscillations and quicker decay. 
The shape function or the final velocity distribution $u(x,y,t)$ is very similar which was presented in Fig.(2) and Fig.(3). 
If the first parameter of Kummer's M function $(\rho /2 + \sigma B_0^2)$ is larger 
than one then the larger the magnetic induction the larger the number of oscillations. This is true for the density and for the 
electric conductivity as well. Such shape functions are presented in 
fig (\ref{valamelyik}). 

%%%%%%%%%%%%%%
% Fig B_dependence N = 1 case 
%%%%%%%%%%%%%%%%%
The third case is $n = 0 $  which is the simplest one.
The original ODE of Eq. (\ref{nine}) is reduced to 
\eq
\eta f' +\sigma B_0^2 f = 0, 
\eqe
with the trivial solution of: 
\eq
f(\eta) = C_1 \eta^{-\sigma B_0^2} . 
\eqe
The velocity distribution has the form
\eq
u(x,y,t) = C_1  \left( \frac{(x+y)}{t} \right)^{-\sigma B_0^2}. 
\eqe  
The pressure field is a pure constant as well. 
These are simple power laws, therefore, we skip to present additional figures.  

\begin{figure}  
\scalebox{0.8}{
\rotatebox{0}{\includegraphics{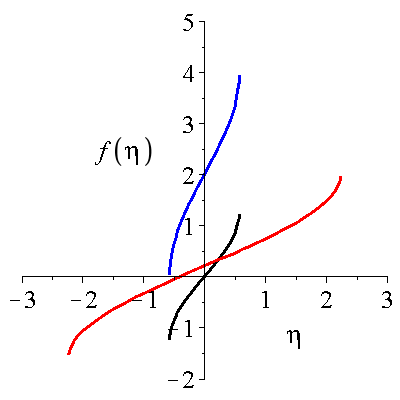}}}
\caption{The velocity shape functions of Eq. (\ref{f9}) for three different parameter sets  $(C_1, C_2, m, \rho )$ with $c_1 =0$. 
The black, red and blue curves are for 
 $(1, 0, 1, 1), (15, 0.2, 2, 1 )$  and $(0.4, 2, 1, 0.4 ) $, respectively}. 
\label{hetes}      
\end{figure}

\begin{figure}  
\scalebox{0.8}{
\rotatebox{0}{\includegraphics{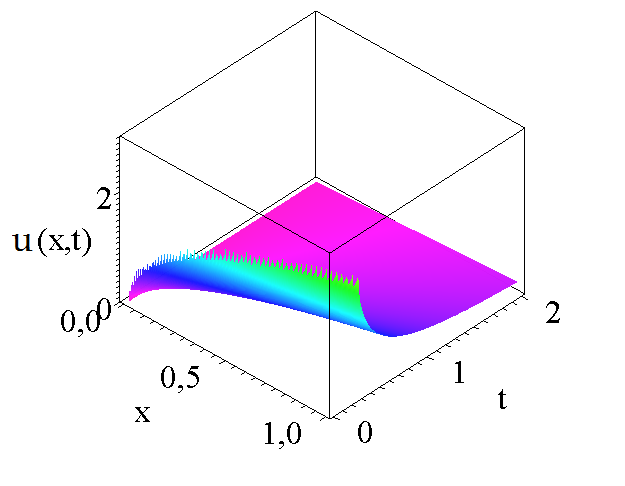}}}
\caption{The velocity distribution which corresponds to  (\ref{f9}) for the parameter set   $( C_1, C_2, \rho,m)$ is  
   $(0, 0, 5, 3)$ }. 
\label{nyolcas}      
\end{figure}

The last case, $n = -1$, is again the most complicated one. 
There is no analytic solution available when all parameters are arbitrary. 
There is however a special case when the last term in Eq.  (\ref{nine}) is zero which defines the constraint of   
$ \rho = 2\sigma B_0^2 $. If 
we additionally fix $c_1 = 0$ we get the following solution: 
\eq
f(\eta) = \pm  \frac{  {1}   }{3 } \sqrt{\frac{6m}{3\rho}}\cdot arctan \left( \frac{\sqrt{6m \rho} \eta}{\sqrt{- 6m \rho  \eta^2 + 2mC_1}}\right)+ C_2.
\label{f9}
\eqe
It can be easily shown that the solution has a compact support, 
and the function is only defined in the region of  $ -\sqrt{\frac{C_1}{3\rho}} \le  \eta \le \sqrt{\frac{C_1}{3\rho}} $. 
Note, that the smaller the fluid density the larger the 
acceptable velocity range if $c_1$ remains the same.  The larger the integral 
constant $c_1$ parameter the larger the available velocity range. The second integral constant $C_2$ 
just shifts the solution parallel to the $y$ axis. 
Figure (\ref{hetes}) shows the solution for three different parameter sets. 
Figure (\ref{nyolcas}) presents the final typical velocity distribution  
$v(x,y=0,t) \sim  t^{\frac{1}{2}}f(\frac{x}{t^{3/2}}) $ the quick temporal decay is clear to see. 
Note, that the prefactor $t^{\frac{1}{2}}$ makes the time backpropagation impossible for 
negative values. 

%%%%%%%%%%%%%%%%%%
% Fig 7
%%%%%%%%%%%%%%%%%%%
% Fig 8
%%%%%%%%%%%%%%%%%%%

For the sake of completeness, we give the solutions for the pressure as well. 
The ODE of the shape function is trivial with the solution of: 
\eq
h' = 0, \hspace*{1cm} h = c_2.
\label{press}
\eqe
Therefore, the final pressure distribution reads: 
\eq
p(x,y,t) = t^{-\gamma} \cdot h(x,y,t) = \frac{c_2}{t^n}, 
\eqe
which means that the pressure is constant in the entire space at a given time, 
and the time decay can be different to the velocity field. 
%%%%%%%%%%%%%%%%%%%%%%%%%%%%%%%%%%%%%%%%%%%%%%%%%%
\subsection{Casson fluid}
To perform an even more comprehensive analysis to study how the transients 
happen in non-Newtonian boundary layers we investigated additional fluid models.  

The first in our line is the simplest one, the so-called Casson fluid model, now 
the boundary layer equation has the form of 
\eq
\rho\left( \frac{\partial u}{\partial  t} +   u\frac{\partial u}{\partial x} + v\frac{\partial u}{\partial y} \right) = - \frac{\partial p}{\partial  x} 
+\left(1+ \frac{1}{\lambda} \right) \mu \frac{\partial^2 u }{ \partial x^2}  + \sigma B_0^2 u,  
\eqe
where $\lambda$ is the Casson parameter. Note that for $\lambda \rightarrow \infty$ 
the term goes over the regular Newtonian viscous term. (All five physical parameters $\rho,\lambda,\mu,B_0 $ and $\sigma$ should 
should have positive real values and the $\lambda \ne 0$ is also evident.) 
The variational approach for the flow of  Casson fluids in pipes was analyzed 
by Sochi \cite{sochi}. The blood flow through a stenotic tube was modelled 
by the Casson fluid by Tandon {\it{et al.}} \cite{tandon}.

We still consider the self-similar Ansatz of (\ref{ansatz})  for the three dynamical variables. 
After the usual algebraic manipulations, we arrive at the non-linear second-order ODE for 
the shape function of horizontal velocity component $f(\eta)$ in the next form:  
\eq
\rho\left(-\frac{f}{2} - \frac{\eta f'}{2}   \right) = \left(1+ \frac{1}{\lambda} \right) \mu  f'' + \sigma B_0^2 f, 
\eqe 
in this model, all four self-similar exponents have fixed values, 
$\alpha = \beta = \delta = 1/2, \gamma = 1$, which is usual for the Newtonian viscous fluid equations \cite{imre4}. 
The derived solutions are mainly different to the first model, therefore we have to give a detailed analysis in the following. 
With our usual mathematical program package Maple 12, we can easily derive the solution 
in closed form:
  \begin{eqnarray}
f(\eta) = e^{-\frac{\rho \eta^2}{4 \mu \left( \frac{\lambda +1}{\lambda} \right)}} \cdot  
\eta \cdot \left( C_1 M \left[- \frac{ 2\sigma B_0^2 -\rho }{2 \rho} ,\frac{3}{2}, \frac{\rho \eta^2}{ 4 \mu \left\{ \frac{\lambda +1}{\lambda} \right\}}  \right] \right. + \nonumber \\ 
 \left.  C_2  U \left[- \frac{ 2\sigma B_0^2 -\rho}{2 \rho} ,\frac{3}{2}, \frac{\rho \eta^2}{ 4 \mu \left\{ \frac{\lambda +1}{\lambda} \right\}}    \right] \right). 
\label{cass}
  \end{eqnarray}  
It is important to note at this point, that the derived formula shows some similarities to the results of the regular diffusion equations \cite{laci_roman} which was exhaustively discussed in our former studies.
The situation is however a bit different here. Thanks to the positivity of all parameters, the sign of exponent is always negative
(which is a real Gaussian function) which dictates a rapid decay for large arguments $\eta$ and for any kind of additional parameter sets. 
The crucial parameter which qualitatively classifies the solutions is the numerical value of the first parameter of Kummer's functions. We 
can distinguish three cases:  
\begin{itemize}
\item{$- \left( \frac{ 2\sigma B_0^2 -\rho }{2 \rho} \right) < 0 $  which is equivalent to $2\sigma B_0^2 > \rho $ (with the stipulation of $\rho \ne 0$)  the solutions have oscillations.  If this parameter is a negative integer the infinite series of the Kummer's function breaks down to a finite order polynomial.  The smaller the parameter the larger the number of oscillations. } 
\item{$ - \left(\frac{ 2\sigma B_0^2 -\rho }{2 \rho} \right) = 0$ which is equivalent to $2\sigma B_0^2 = \rho $ (with the stipulation of $\rho \ne 0$)  now both Kummer's M and Kummer's U functions are unity and 
the solution is reduced to the Gaussian function times $\eta $ function which has odd symmetry. This is the limiting solution 
between the oscillating and the non-oscillating solutions. }
\item{$- \left( \frac{ 2\sigma B_0^2 -\rho }{2 \rho} \right) > 0$  which is equivalent to $2\sigma B_0^2 < \rho $ (with the stipulation of $\rho \ne 0$)  the larger the parameter the smaller peak value of the solution and the quicker the decay.   }
\end{itemize} 

\begin{figure}  
\scalebox{0.4}{
\rotatebox{0}{\includegraphics{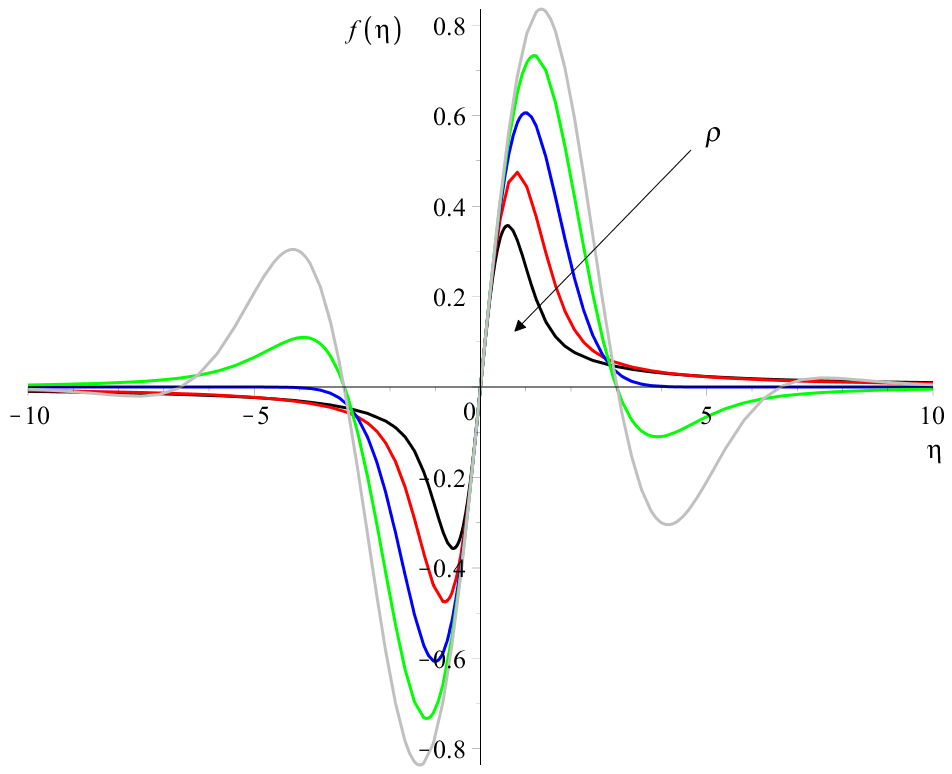}}}
\caption{Five shape functions of Eq. (\ref{cass}) for different parameter sets. The role of the fluid density $\rho$ is investigated. 
The black, red, blue green and grey lines are for $(C_1, C_2, \rho,\lambda, \mu, \sigma, B_0^2) $ with 
numerical values of $ (1,0,8,1,1,1,1), (1,0,4,1,1,1,1), (1,0,2,1,1,1,1), (1,0,1,1,1,1,1)$ and $ (1,0,1/2,1,1,1,1)$, respectively}. 
\label{kilences}      
\end{figure}

Figure (\ref{kilences}) presents five shape functions  of Eq. (\ref{cass}) for five different fluid densities, for a clear comparison all the other parameters remain the same. The larger the density $\rho$ the quicker the 
decay and the smaller the global maximum  
of the shape functions. 

%%%%%%%%
% Fig 9

%%%%%%%%%%%%%%%%%
% Fig 10b
\begin{figure}  
\scalebox{0.4}{
\rotatebox{0}{\includegraphics{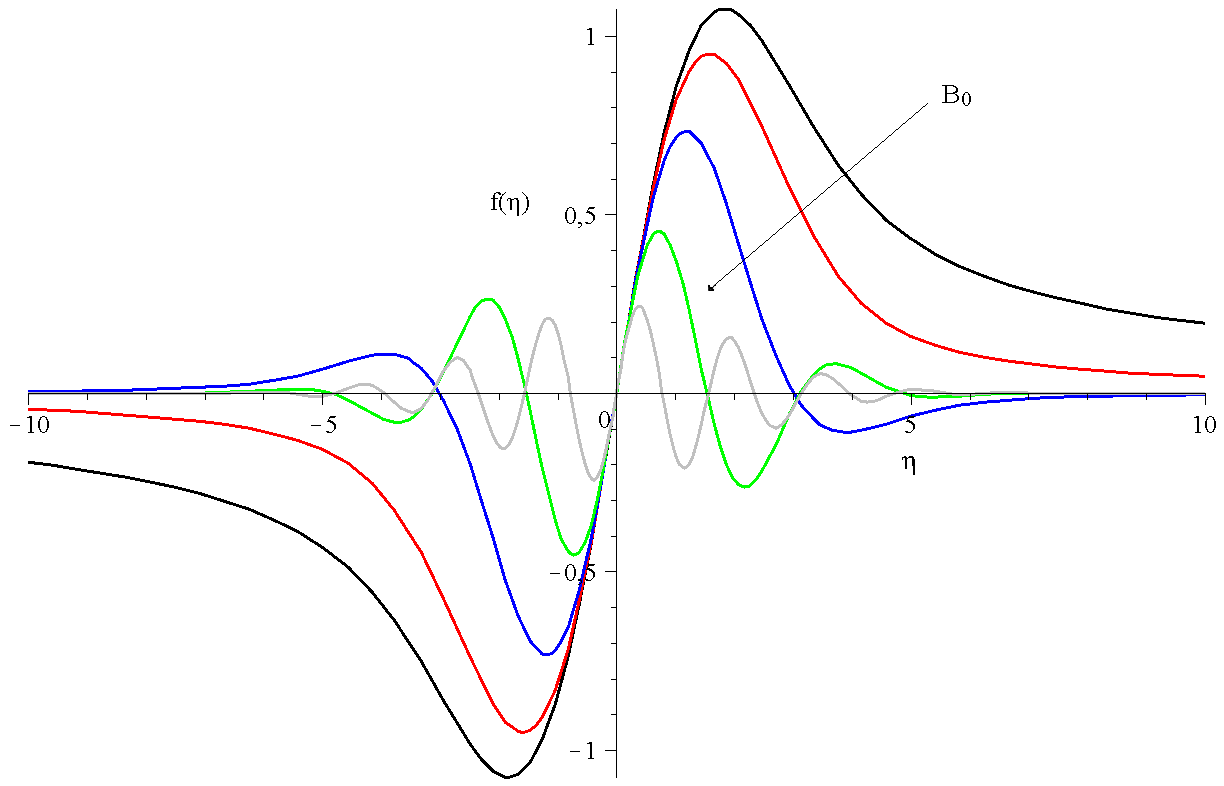}}}
\caption{Five shape functions of Eq. (\ref{cass}) for different parameter sets.   The role of the magnetic induction $B_0^2$is investigated. 
The black, red, blue  and green lines are for $(C_1, C_2,\rho,\lambda, \mu, \sigma, B_0^2) $ with 
numerical values of $ (1,0,1,1,1,1,0.1), (1,0,1,1,1,1,0.5), (1,0,1,1,1,1,1), (1,0,1,1,1,1,2)$ and $ (1,0,1,1,1,1,4)$ respectively}. 
\label{tizes_b}      
\end{figure}
%%%%%%%%%%%%%%%%%

In general, the ${\rho}/[{4\mu (\lambda +1)/\lambda]} $ factor is responsible for the extent  (or the full with at half maximum (FWHM) value
 of the solution) the larger this parameter the quicker the decay.  
For changing the Casson parameter in the physically relevant positive range $\lambda > 0$  we found no remarkable change in the solutions,  
this is because the  $\frac{\lambda +1}{\lambda}$   (is almost unit) factor appears in the Gaussian together with two other parameters. 

The role of the magnetic induction is also important to know. 
Figure (\ref{tizes_b}) shows the effect of the magnetic induction $B_0^2$  
where all the other parameters and unchanged. 
If the first parameter of Kummer's M function is smaller than unity the solutions 
has a global maxima and a quick decay, if this parameter is larger than the unity 
then the larger the magnetic induction the larger the oscillations of the solutions.  

Most of these solutions have odd symmetry. 
It can be easily seen because the Gaussian is 
an even function $\eta$ is an odd function and the series of Kummer's M and U functions which have a quadratic argument   
are again even infinite or finite series which together given odd 
symmetry. 
However, using the series expansion it can be shown that if the first 
parameter of the Kummer's U (only for U ) function
$-\left( \frac{2\sigma B_0^2 -\rho }{2 \rho}\right)  $  is a negative half-integer then the solutions get even symmetry. 
The properties of such kinds of solutions are exhaustively discussed in \cite{laci}. 

To complete our analysis we give the solutions for the pressure as well. 
The ODE of the shape function is trivial with the solution of: 
\eq
h' = 0, \hspace*{1cm} h = c_2.
\label{press}
\eqe
Therefore, the final pressure distribution is: 
\eq
p(x,y,t) = t^{-1} \cdot h(x,y,t) = \frac{c_2}{t^{-1}}, 
\eqe
which means that the pressure is constant in the entire space at a given time, and the temporal decay 
follows the simple inverse law. 

%%%%%%%%%%%%%%%%%%%%%%%%%%%
\subsection{Oldroyd-B model}
The next fluid flow model is the Oldroyd-B model \cite{old}, where the momentum equation is the following:
  \begin{eqnarray}
\frac{\partial u}{\partial t}+ u \frac{\partial u}{\partial x} + v \frac{\partial v}{\partial y} 
+ \frac{\lambda_1}{\rho} \left( u^2 \frac{\partial^2 u}{\partial x^2} + 2 u v \frac{\partial^2 u}{\partial x \partial y}  +   
v^2 \frac{\partial^2 u}{\partial y^2}\right) = -\frac{1}{\rho}\frac{\partial p}{\partial x} + 
\frac{\nu}{\rho} \frac{\partial^2 u}{\partial y^2} 
\nonumber \\
+ \frac{\nu \lambda_2}{\rho} \left( u \frac{\partial^3 u}{\partial x \partial y^2} +  
v \frac{\partial^3 u}{\partial y^3}  -  \frac{\partial u}{\partial y}\frac{\partial^2 u}{\partial y^2}
- \frac{\partial u}{\partial y}\frac{\partial^2 v}{\partial y^2}\right)+\sigma B_0^2 u .
  \end{eqnarray}

Even without the external term with the magnetic induction, the self-similar Ansatz of (\ref{ansatz})  
leads to contradiction among the four self-similar exponents indicating that the system has no self-similar 
symmetry with power-law time decay. The main cause of the lack of symmetry is the appearance of the third 
spatial derivatives together with the second ones. 
 The Oldroyd-B model was exhaustively investigated in the last decades from many points of 
 views, optimal time decay rates for the higher order spatial derivatives of solutions were analysed by Wang \cite{wang}. 
 The global wellposedness of the model was investigated by Elgindi and Liu 
 \cite{elg}. The global existence results of some Oldroyd-B models were proven by \cite{hieber}. Finally, we mention the review paper of 
 \cite{rev} which summarizes most of the known mathematical results. 
 
%%%%%%%%%%%%%%%%%%%%%
\subsection{Walters' Liquid B model}
The third possible non-Newtonian candidate is the so-called Walters' Liquid B model, \cite{walt}  where the 
momentum equation has the form of: 
\begin{equation} 
 \frac{\partial u}{\partial t} + 
u \frac{\partial u}{\partial x} + v \frac{\partial v}{\partial y} =  -\frac{1}{\rho}\frac{\partial p}{\partial x} + \frac{\mu}{\rho} \frac{\partial^2 u}{\partial y^2} + \frac{k_0}{\rho} \left( u \frac{\partial^3 u}{\partial x \partial y^2} + v \frac{\partial^3 u}{\partial y^3} - \frac{\partial u}{\partial y}\frac{\partial^2 u}{\partial x \partial y} + \frac{\partial u}{\partial x}\frac{\partial^2 u}{\partial y^2}\right)+\sigma B_0^2 u.    
\end{equation}
Practically, we observed the same as in the previous case, the existence of the terms with the third derivatives 
destroyed the self-similar symmetry therefore no solutions can be derived.  

Flow and heat transfer of Walter's liquid model with stretching walls 
were applied to haemodynamics by Misra, Shit and Rath \cite{mis}. 
Khrisna \cite{kris} studied the 
   Hall and ion slip effects on MHD laminar flow of in a Walter's‐B fluid. 
   
%%%%%%%%%%%%%%%%%%%%%%%
\subsection{Williamson fluid }
In the last  model, we took the Williamson fluid \cite{wil}
 with the  equation of motion of: 
\begin{equation}
\frac{\partial u}{\partial t} + 
u \frac{\partial u}{\partial x} + v \frac{\partial v}{\partial y}   = - \frac{1}{\rho} \frac{\partial p}{\partial x} + \frac{\partial }{\partial y} \left( \frac{\mu}{\rho} \frac{\partial u}{\partial y} + \mu  \frac{\Gamma}{2} \left(\frac{\partial u}{\partial y} \right)^2  \right)+\sigma B_0^2 u.     
\end{equation}
Even if this model lacks the property of time-dependent self-similar symmetry, therefore we cannot present analytical results. %As we saw in the literature even this 
%model is under heavy investigations. 
%the time-independent versions of all these last three models are under heavy investigations 
%with different reduction techniques and symmetry considerations having very nice results. 
%As an example we mention that 
Malik, et al. \cite{malik}  however numerically studied the effects of variable thermal conductivity and heat generation/absorption on Williamson fluid flow and heat transfer.
The swimming effects of microbes in blood flow of nano-bioconvective Williamson's fluid was investigated by Rana {\it{et all.}} \cite{rana}.
We found these recent results which we think are worth mentioning. 
%%%%%%%%%%%%%%%%%%%%%%%%%%%%%%
\section{Summary and Outlook} 
In our study, we investigated five different time-dependent non-Newtonian boundary 
layer problems using the self-similar approach. We show that analytical results exist and give them. 
For the non-Newtonian power-law fluid, we found different types of solutions depending 
on the value of the power-law exponent, which can be expressed by Kummer functions or 
other implicit functions. The effects of density and other parameters in the nonlinear 
ordinary differential equation have been investigated.
The effects of the magnetic field can be described by including an additional magnetic 
term in the equation. Analytical solutions have also been found for such cases.
For non-Newtonian Casson fluid flow, the effect of the Casson parameter is also analyzed.
For the other three non-Newtonian cases, the Oldroyd-B model, Walter's Liquid B model, and Williamson 
fluid, the self-similar transformation (\ref{ansatz}) cannot be performed, for these cases the solution according 
to the power law and the Casson fluid cannot be given.
All five models might be investigated with the travelling wave Ansatz in the far future because 
the spatial and temporal symmetry shift is always present in these equations. 

%%%%%%%%%%%%%%%%%%t%%%%%%%%%%%%%%%
\section{Acknowledgments}
One of us (I.F. Barna) was supported by the NKFIH, the Hungarian National Research Development and Innovation Office. 
This work was supported by project no. 129257 implemented with the
support provided by the National Research, Development and
Innovation Fund of Hungary, financed under the $K \_ 18$ funding scheme. \\
Author Contributions:  I.F.B. Conceptualization, analytic calculations, manuscript 
writing; G.B. Conceptualization, Literature;   K.H.  Literature collection, 
writing the final manuscript; L.M. Correction of the manuscript \\
Data Availability Statement: The data that supports the findings of this study are all available within the article. \\
Conflicts of Interest: The authors declare no conflict of interest.

%%%%%%%%%%%%%%%%%%%%%%%%%%%%%%%%%%%%%%%%
 

\begin{references}
 
\bibitem{arista} G. Astarita and G. Marucci. {\it{Principles of non-Newtonian fluid mechanics}}, 
McGraw-Hill, London, New York, 1974.

\bibitem{non-newt} I. Fridtjov   {\it{Rheology and Non-Newtonian Fluids}},  Springer, 2014. 

\bibitem{patel}  M. Patel and M. Timol, {\it{Non-Newtonian fluid models and Boundary Layer flow }},   LAP LAMBERT Academic Publishing, 2020. 

\bibitem{schli}H. Schlichting and K. Gersten  {\it{Boundary-layer theory}} 
Berlin Heidelberg New York: Springer; 2016.

\bibitem{yori} Y. Hori {\it{Hydrodynamic lubrication}} Tokyo: Springer; 2006. 
%https://doi. org/ 10. 1007/4- 431- 27901-6_2. 

\bibitem{barna-bound} I. F. Barna, G. Bogn\'ar,  L. M\'aty\'as and K. Hricz\'o, 
Journal of Thermal Analysis and Calorimetry  {\bf{147}}, (2022) 13625.
%–13632
%https://doi.org/10.1007/s10973-022-11574-3

\bibitem{saer} C. Saengow and  A.J. Giacomin,  
%Exact solutions for oscillatory shear sweep behaviors of complex fluids from the Oldroyd 8-constant framework. 
Phys Fluids. {\bf{30}}, (2018) 030703. 
%https:// doi. org/ 10.1063/1. 50235 86.

\bibitem{bog}G.  Bogn\'ar,
%Similarity solution of boundary layer flows for nonnewtonian fluids. 
Int J Nonlinear Sci Numer Simul. {\bf{10}}, (2009) 1555. 
%;10(11–12):1555–66. 
% https:// doi. org/ 10. 1515/ IJNSNS. 2009. 10. 11- 12.1555.

\bibitem{bog2} G. Bogn\'ar and K. Hricz\'o,  
%Similarity solution to a thermal boundary
%layer model of a non-Newtonian fluid with a convective
%surface boundary condition. 
Acta Polytechnica Hungarica, {\bf{8}}, (2011) 131.
%;8(6):131–40.

\bibitem{mak} T. M. Ajayi, A. J. Omowaye, and I. L. Animasaun,  
Journal of Applied Mathematics  2017, Article ID 1697135.  
  %%
\bibitem{casson} N. Casson, {\it{ Rheology of Disperse Systems}} 
(Edited by Mill, D. C.) pp. 84--102. Pergamon Press, Oxford, 1959. 

\bibitem{old} J. Oldroyd, 
%On the Formulation of Rheological Equations of State. 
Proc. Roy. Soc. London Series A, Math. and Phys. Sciences 
{\bf{200}}, (1950) 523.
%–541  %Bibcode:1950RSPSA.200..523O. doi:10.1098/rspa.1950.0035.

\bibitem{walt} 	K. Walters,
%Non -Newtonian effects in some elastic-viscous liquids whose behavior at small rates of shear is characterized by a general linear equations of state, 
Quart. J. Mech. Appl. Math.  {\bf{6}}, (1963) 63.

\bibitem{wil}R.V.  Williamson,
% The flow of pseudoplastic materials. 
Ind. Eng. Chem. {\bf{21}}, (1929) 1108.

\bibitem{power1} S. Ko,
%Temporal decay of strong solutions for generalized Newtonian fluids with variable power-law index
J. Math. Phys. {\bf{63}}, (2022) 041508. 

\bibitem{power2}R.S. Herbst, C. Harley, K.R. Rajagopal, 
%Flow reversals of a non-Newtonian fluid in an expanding channel
International Journal of Non-Linear Mechanics, 
{\bf{154}}, (2023) 104445. 
%DOI:10.1016/j.ijnonlinmec.2023.104445

\bibitem{power3} R. Fazio, 
%A non-iterative transformation method for boundary-layer with power-law viscosity for non-Newtonian fluids
Calcolo, {\bf{59}},  (2022)  43. 

\bibitem{patel} M. Patel, H. Surati and M. G. Timol, Math. J. Interdiscip. {\bf{9}}, (2021) 35.
%–41

\bibitem{sedov} L. Sedov, {\it{Similarity and Dimensional Methods in Mechanics}}, CRC Press, 1993.

%\bibitem{imre1} I.F. Barna and L. M\'aty\'as, Chaos Solitons and Fractals {\bf{78,}}  249 (2015). 

\bibitem{imre1} I.F. Barna, M.A. Pocsai, S. L\"ok\"os and L. M\'aty\'as, Chaos Solitons and Fractals {\bf 103}, (2017) 336.

\bibitem{imre3} I.F. Barna,  L. M\'aty\'as and M.A. Pocsai, Fluid. Dyn. Res. {\bf{52}}, (2020) 015515.

\bibitem{imre4} I.F. Barna, 	Commun. Theor. Phys. {\bf{56}}, (2011) 745.

 \bibitem{NIST}F. W. J. Olver, D. W. Lozier, R. F. Boisvert and C. W. Clark
{\it{NIST Handbook of Mathematical Functions}} Cambridge University Press, 2010. 

\bibitem{meta}F. Cappolino {\it{Theory and Phenomena of Metamaterials}} 
CRP Press, 2009.

\bibitem{laci_roman}  L. M\'aty\'as and I.F. Barna
%General Self-Similar Solutions of Diffusion Equation and Related Constructions,
Romanian Journal of Physics {\bf{67}}, (2022) 101.  

\bibitem{laci} L. M\'aty\'as and I.F. Barna, 
%"Even and Odd Self-Similar Solutions of the Diffusion Equation for Infinite Horizon"
Universe {\bf{9}}, (2023) 264.  

\bibitem{sochi} T. Sochi,  	International Journal of Modeling, Simulation, 
and Scientific Computing, {\bf{7}}, (2016) 1650007. 

\bibitem{tandon} P.N. Tandon, U.V.S. Rana, M. Kawahara, V.K. Katiyar, 
% A model for blood flow through a stenotic tube,
International Journal of Bio-Medical Computing,  {\bf{32}},
(1993) 61.
%-78,

\bibitem{wang}  Y. Wang,
Z. Angew. Math. Phys. {\bf{74}},  (2023) 3. 

\bibitem{elg} T.M. Elgindi, J.L. Liu, 
%Global wellposedness to the generalized Oldroyd type models in R3. J.
Differential Equations, {\bf{259}}, (2015) 1958
% –1966

\bibitem{hieber}  M. Hieber, Y. Naito, Y. Shibata, 
%Global existence results for Oldroyd-B fluids in exterior domains.
J. Differential Equations, {\bf{252}}, (2012) 2617.
%–2629

\bibitem{rev}M. Renardy, B. Thomases, 
%A mathematician’s perspective on the Oldroyd B model: progress and
%future challenges. 
J. Non-Newton. Fluid Mech., {\bf{293}}, (2021)  104573.
%, 12 pp

\bibitem{mis}  
J. C. Misra, G. C. Shit and H. J. Rath, 
% Flow and Heat Transfer of a MHD Viscoelastic
% Fluid in a Channel with Stretching Walls:
% Some Applications to Haemodynamics
Computers \& Fluids, {\bf{37}},  (2008) 1.
%-11

\bibitem{kris} M. Veera Krishna,
%Hall and ion slip effects on MHD laminar flow of an elastico-viscous (Walter's-B) fluid
Heat Transfer {\bf{49}}, (2020) 2311.
%-2329

\bibitem{malik} M.Y. Malik, M. Bibi, F. Khan and T. Salahuddin, 
%Numerical solution of Williamson
%fluid flow past a stretching cylinder and heat transfer with variable thermal
%conductivity and heat generation/absorption, 
AIP Advances, {\bf{6}}, (2016) 035101.

\bibitem{rana} B.M.J. Rana, S.M. Arifuzzaman, Saiful Islam, Sk. Reza-E-Rabbi, 
Abdullah Al-Mamun, Malati Mazumder, Kanak Chandra Roy, Md. Shakhaoath Khan, 
Thermal Science and Engineering Progress,
{\bf{25}}, (2021) 101018.

\end{references}
\end{document}